\documentclass[12pt]{article}

\usepackage[margin=1in]{geometry}
\usepackage[english]{babel}

\usepackage[numbers]{natbib}
\usepackage{authblk}

\usepackage{placeins}
\usepackage{amssymb,amsmath,empheq,xcolor}
\usepackage{wasysym}
\usepackage{placeins} 
\usepackage{xcolor} 
\usepackage{xspace} 

\usepackage{mathtools}
\usepackage{siunitx}
\usepackage{diffcoeff}
\usepackage{upgreek}
\usepackage[normalem]{ulem}
\usepackage[utf8]{inputenc}
\usepackage{float}
\usepackage[normalem]{ulem}

\usepackage{lineno}
\usepackage{graphicx}

\usepackage{tikz}
\usetikzlibrary{arrows.meta,positioning,fit,calc}

\usepackage[pdfencoding=auto]{hyperref}
\usepackage{hyperref}
\usepackage{svg}
\usepackage{comment}

\title{Compact VMM3a/SRS integration for high-channel-density TPC readout}

\author[1]{Alexandru Enciu}
\author[1]{Uwe Bonnes}
\author[1,2]{Meytal Duer}
\author[1,2,3]{Piotr Gasik}
\author[1]{Andrea Lagni}
\author[2]{Bastian L\"oher}
\author[1]{Leandro Milhomens de Fonseca}
\author[1,4]{Alexandre Obertelli}
\author[2]{Martin Poghosyan}
\author[5]{Hans T\"ornqvist}
\author[6]{Yanzhao Wang}
\author[1]{Frank Wienholtz}

\affil[1]{Technische Universit\"at Darmstadt, Fachbereich Physik, Darmstadt, Germany}
\affil[2]{GSI Helmholtzzentrum f\"ur Schwerionenforschung GmbH, Darmstadt, Germany}
\affil[3]{FAIR Facility for Antiproton and Ion Research in Europe GmbH, Darmstadt, Germany}
\affil[4]{Helmholtz Forschungsakademie Hessen f\"ur FAIR (HFHF), Frankfurt, Germany}
\affil[5]{Department of Physics, Chalmers University of Technology, G\"oteborg, Sweden}
\affil[6]{Institut f\"ur Kernphysik, Universit\"at zu K\"oln, K\"oln, Germany}

\date{}

\begin{document}
\maketitle

\begin{abstract}
The HYDRA (HYpernuclei Decay at R$^3$B Apparatus) TPC is a 5,376-channel pion tracker developed for hypernuclear studies at the R$^3$B experiment at GSI/FAIR. Its installation inside the GLAD dipole magnet imposes strict space constraints above the pad plane. To fit the front-end electronics within this volume, the VMM3a/SRS readout was adapted through a vertical hybrid arrangement and custom boards for signal routing and power distribution. At the same time, the readout had to be integrated with the R$^3$B timing system, which is not natively supported by the SRS. This was achieved by recording the R$^3$B heimtime and trigger signals through spare VMM3a channels, allowing the global timestamps to be reconstructed from the same data stream. Simulations of the PCB coupling indicated low crosstalk between adjacent traces, while S-curve measurements gave a baseline noise level consistent with that reported for the standard VMM3a/SRS readout. Cosmic-muon measurements with the plastic wall provided reconstructed tracks and confirmed event-by-event timing correlation between the two detector systems.
\end{abstract}

\section{Introduction}

A time projection chamber enables the three-dimensional reconstruction of charged-particle trajectories \cite{Sauli_2023}. The positions of the signals induced on the segmented pad plane provide the two transverse coordinates, while the electron drift time determines the longitudinal coordinate. TPCs are widely used in nuclear and particle physics experiments and are often operated inside magnetic fields, where the curvature of a particle trajectory provides information about its momentum and charge sign. Their configurations range from large cylindrical detectors used in collider experiments, such as the ALICE TPC at CERN \cite{Alme2010}, to compact active-target systems in which the detector gas also serves as the reaction target, such as the AT-TPC at NSCL-FRIB \cite{Bradt2017}.

\begin{figure}[t]
\includegraphics[width=\textwidth]{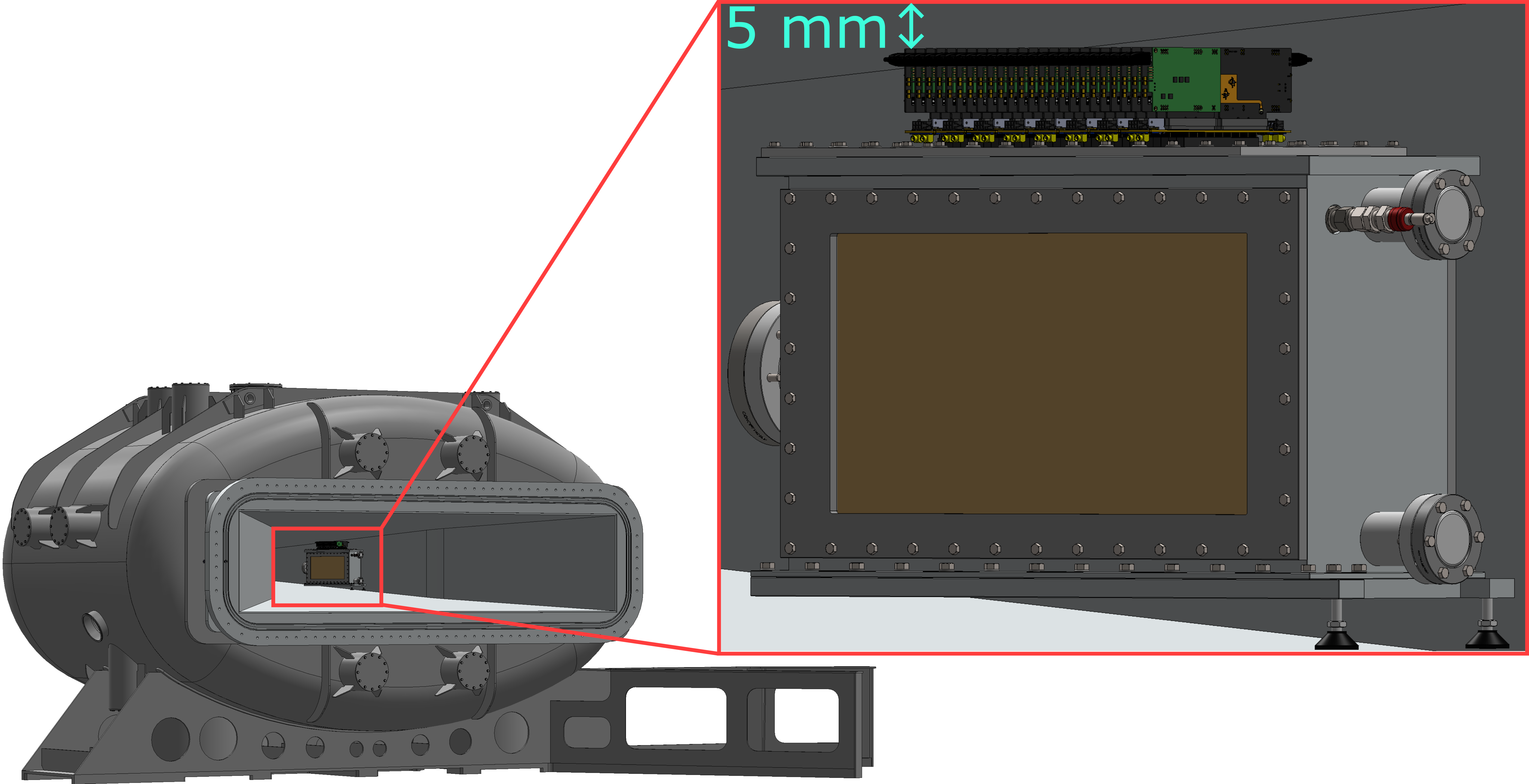}
\centering
\caption{HYDRA TPC and VMM3 readout inside the GLAD magnet; the right zoom-in highlights the space constraints between the last readout VMM3 card and the magnet ceiling.}
\label{figure: SRS_System}
\end{figure}

The HYDRA tracker is a compact TPC developed as a pion tracker for hypernuclear studies at the Reactions with Relativistic Radioactive Beams (R$^3$B) experiment at GSI/FAIR. It has a drift length of 300 mm and uses a Gas Electron Multiplier (GEM) \cite{Sauli1997}, followed by a Micro-Mesh Gaseous Structure (Micromegas) \cite{Giomataris1996}, for charge amplification. The amplified signals are induced on a pad plane comprising 5,376 instrumented pads, each measuring $1.9 \times 1.9$~mm$^2$. A plastic scintillator wall provides the trigger and event-time reference required to determine the electron drift time. HYDRA is designed to operate inside the large-acceptance GLAD dipole magnet at R$^3$B \cite{Gastineau2008} (see Fig.~\ref{figure: SRS_System}), allowing the trajectories and momenta of charged particles to be reconstructed. A detailed description of the HYDRA tracker is given in Ref.~\cite{Ji2026}, while its scientific program is discussed in Ref.~\cite{Velardita2023}.

The hypernuclear production cross sections relevant to HYDRA are of the order of $\mathcal{O}(100~\upmu\mathrm{b})$. Consequently, the detector must operate at high beam intensities and sustain correspondingly high particle rates. Its readout must therefore provide continuous high-rate operation together with charge and timing measurements for each recorded signal. The VMM3a front-end ASIC was developed specifically for the readout of gaseous detectors and provides a time resolution of $\mathcal{O}(1~\mathrm{ns})$ while supporting per-channel rates in the MHz range \cite{Geronimo2022}. Its integration into the Scalable Readout System (SRS) provides a scalable architecture for reading the 5,376-channel HYDRA pad plane \cite{Lupberger2018}.

The standard VMM3a/SRS configuration could not be installed within the limited space available above the HYDRA pad plane inside the GLAD magnet. The VMM Hybrids were therefore mechanically modified and connected to the detector through custom card adapters and pad-plane adapter modules. These modules also provide local power distribution because the SRS crate must be located several meters from the detector. A further requirement was the synchronization of the SRS readout with the R$^3$B data-acquisition system, although the SRS does not natively support an R$^3$B timestamp receiver. To address this limitation, the R$^3$B heimtime and trigger signals are distributed to the detector and AC-coupled into spare VMM3a channels, where their edges are timestamped together with the detector signals.

This paper describes the mechanical and electrical adaptation of the VMM3a/SRS readout for HYDRA and evaluates the resulting system. Coupling from the 3.3~V timing-signal traces into adjacent detector-channel traces is estimated through circuit simulation. The physical noise of the assembled readout is determined independently from threshold S-curves and expressed as equivalent noise charge (ENC). Finally, cosmic-ray measurements performed with the plastic scintillator wall are used to demonstrate track reconstruction and event-by-event timing correlation between the two detector systems.

\section{Adaptation of VMM Hybrids}\label{sec:vmm}

\subsection{VMM Hybrids and SRS readout}\label{sec:VMM_SRS}

The VMM3a is a 64-channel front-end ASIC, initially developed for the ATLAS muon system upgrade at CERN, and used to read out Micromegas and small-strip Thin Gap Chamber detectors \cite{Geronimo2022}. The signal processing for each channel starts with a preamplifier, for which the gain can be adjusted into eight values between 0.5 to 16 mV/fC. This is followed by a shaper with a programmable peaking time of 25, 50, 100, and 200 ns, a discriminator at which the threshold level is set, and a peak finder. The output of the latter provides the charge and time information, which is then digitized. The VMM3a has multiple operating modes, where a continuous readout is used for the HYDRA TPC, allowing rates up to a maximum of 4 MHz per channel without deadtime limitation.

To provide a general-purpose readout system for gaseous detectors, the VMM3a was integrated into the SRS developed by the RD51 collaboration at CERN \cite{Martoiu2013}. For this purpose, the VMM Hybrid \cite{Lupberger2018,Pfeiffer2022} was implemented, a 5$\times$8 cm$^2$ Printed Circuit Board (PCB) that hosts two VMM3a chips, providing 128 channels in total. A 140-pin Hirose FX10A series connector on the hybrid connects to the detector readout. From the Hirose connector, 128 channels are routed to the inputs of the two VMM3a. Other I/Os on the ASICs, such as clock inputs, configurations, and data output, are provided by an AMD Xilinx Spartan-7 series Field Programmable Gate Array (FPGA). Additional components on the Hybrid are low-dropout regulators for a stable power supply, spark protection networks, and connectors for high-speed digital links. The signals are AC-coupled and protected by arrays of transient-voltage-suppression diodes, while on-board coupling networks stabilize each channel's baseline. The Hybrids are equipped with low-impedance ground connections and cooling plates, as heat dissipation reaches approximately 4 W per board.

From the Spartan-7 FPGA, the digitized signals are sent via the Digital VMM (DVMM) adapter to the Front-End Concentrator (FEC) card. The Hybrid is connected via an HDMI cable to the DVMM, where power can be provided over the cable itself for a short HDMI cable (up to 2 m in length). The FEC is built around an Xilinx Virtex-6 FPGA and handles configuration, event buffering, zero suppression, and data transmission to the DAQ PC over a Gigabit Ethernet connection. One FEC can accommodate up to eight Hybrids (1024 channels). With multiple FECs in operation, a Clock and Trigger Fanout (CTF) module distributes a common 44 MHz base clock, ensuring synchronization across the system. The FECs and CTF card are housed and powered by the SRS Power Crate, which can also provide power to the Hybrids via the DVMM. Data from the SRS is grouped in UDP (User Datagram Protocol) frames and sent to the DAQ PC. 

\subsection{Requirements for the HYDRA TPC at R\texorpdfstring{$^3$}{3}B}

The HYDRA TPC pad plane consists of $44 \times 128$ square pads, each with an area of $1.9 \times 1.9$ mm$^2$, separated by a 0.1 mm gap \cite{Ji2026}. Along the 44 rows, the first and last rows are grounded, leaving 5,376 signal pads. The readout of each pad is performed individually via a 2-mm-pitch pin header.

For the target experiments with HYDRA, the high beam intensities $\mathcal{O}$(10$^6$ particles per second) imply about 200 kHz of charged particles traversing the drift region of the TPC. The front-end electronics (FEE) must therefore sustain continuous high-rate operation with minimal dead time to avoid efficiency losses. Equally important is low-noise performance, since the desired spatial resolution of 300 $\upmu$m can only be achieved if the charge deposited on the signal pads is measured with high accuracy and stability. To eliminate the need for additional cabling and reduce the input capacitance, the FEE must be mounted as close as possible to the signal pins of the TPC. As the setup will operate inside the GLAD dipole magnet, the consequent reduced space further constrains the mechanical layout. Finally, the TPC readout must be synchronized with the main R$^3$B DAQ system.

\subsection{Adapter boards and VMM Hybrid modifications}

\begin{figure}[ht]
\includegraphics[width=\textwidth]{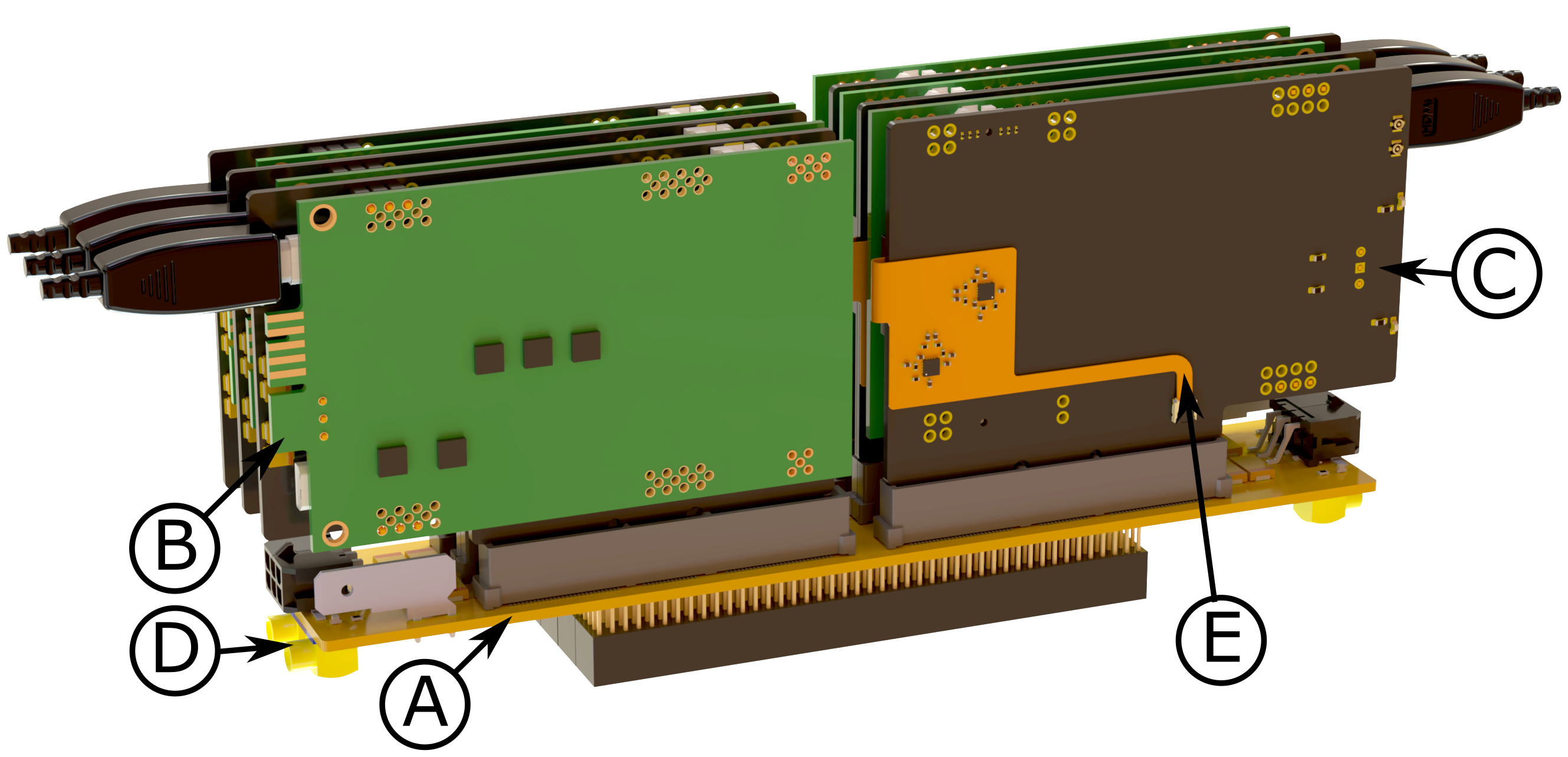}
\centering
\caption{1/8th HYDRA TPC readout: (A) Pad-plane adapter. (B) Modified VMM3 Hybrid. (C) VMM3 Hybrid card adapter PCB. (D) 2-input-to-6-output add-on clock buffer. (E) 2-input to 8-output Flexible PCB add-on clock buffer.}
\label{figure: HYDRA in GLAD}
\end{figure}

For readout of the full HYDRA TPC, a total of 48 VMM Hybrids are used, and with the space limitations on top of the pad plane, modifications of the Hybrids were required, and consequently, an adapter PCB card attached to each Hybrid was designed. To connect the modified Hybrids via the card adapter to the pin headers on the pad plane, an adapter module was implemented, which also provides power to the boards. View of the setup mounted on the TPC pad plane is shown in Fig.~\ref{figure: SRS_System}, with detailed schematics for one readout module in Fig.~\ref{figure: HYDRA in GLAD}. In the following, we detail the specific developments. The schematics, PCB layouts, and bills of materials for the HYDRA TPC readout electronics are archived on Zenodo \cite{enciu_2026_22640763}.

\subsubsection{Pad-plane adapter module}

\begin{figure}[ht]
\includegraphics[width=\textwidth]{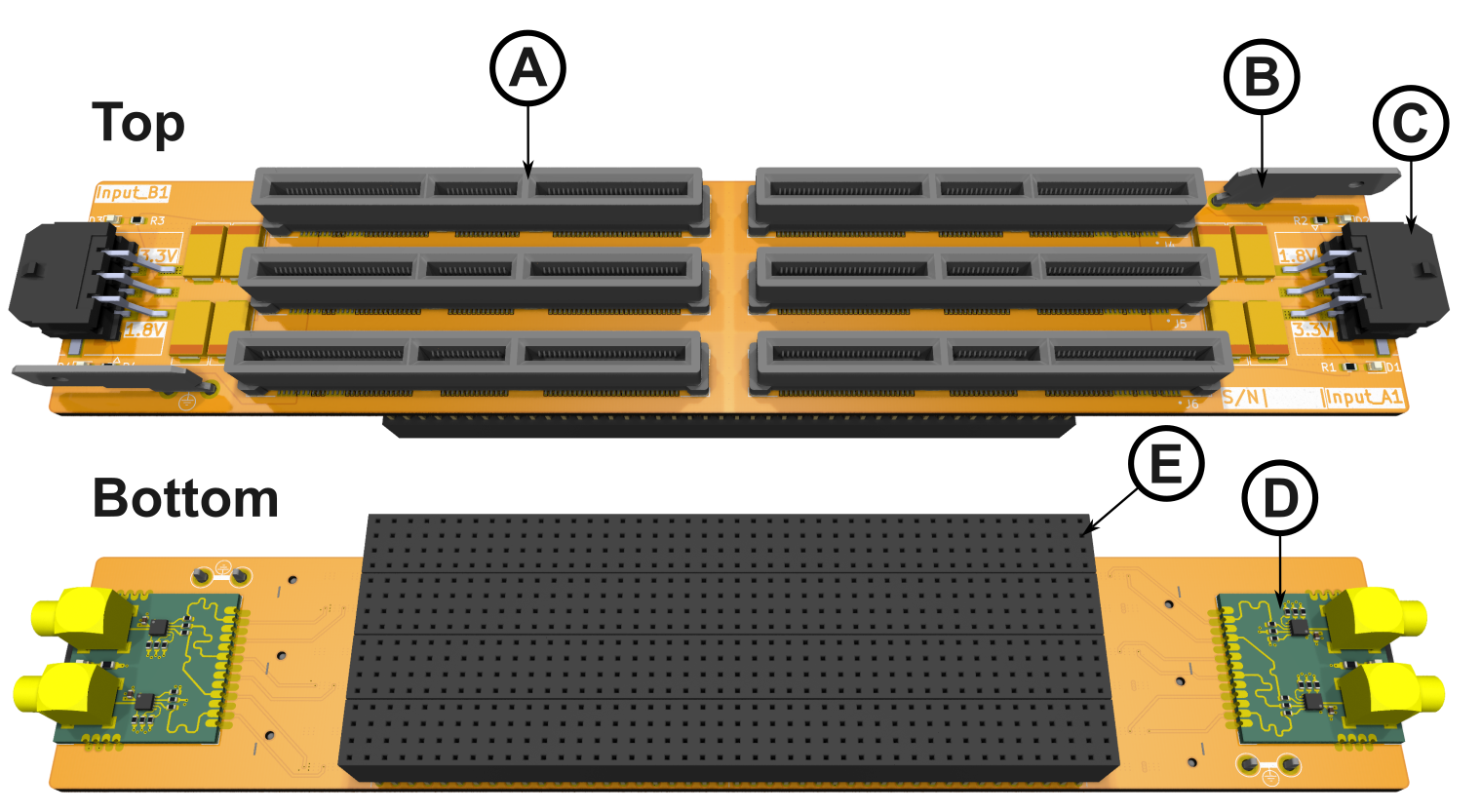}
\centering
\caption{Pad-plane adapter: (A) 140-pin card connector. (B) Fast-On ground connector. (C) Molex Micro-Fit 3.0 6-pin power connector. (D) 2 inputs to 6 outputs add-on clock buffer. (E) 2-mm pitch pin-socket connectors.}
\label{figure: Pad plane adaptor}
\end{figure}

The TPC pad-plane readout is divided into eight identical sections. Each section transfers the signals from 672 pads and routes them to six VMM Hybrids. The first component is an adapter module (see Fig. \ref{figure: Pad plane adaptor}), plugged into the pad plane via a 2-mm pitch Through-Hole Technology (THT) pin sockets and attached to a PCB where the signals towards the VMM3a are routed. The latter is an eight-layer High-Density Interconnect (HDI) PCB implemented with buried and blind vias and 125 $\upmu$m controlled-impedance traces, with each signal layer referenced to a dedicated ground plane. 

The adapter module also provides power to the Hybrids via a robust Molex Micro-Fit 3.0 six-pin connector on both sides of the PCB. As described in Sec. \ref{sec:VMM_SRS}, with short HDMI cables, the power can be provided directly via the cable itself. For the HYDRA application, a cable length of several meters is required as the SRS crate is placed outside the GLAD magnet. In such a case the power loss in the thin HDMI conductors would exceed the power required by the Hybrid. Therefore, the boards have to be powered by an external power supply. The power rails (1.8 V and 3.3 V) are locally decoupled with two 100 $\upmu$F tantalum polymer capacitors to ensure stable supply conditions. If additional grounding is required, a Fast-On ground connector is available on the board for reliable, modular installation.

A custom add-on clock buffer is implemented on each side of the adapter module to distribute synchronization signals (see Sec. \ref{sec:sync}) to each Hybrid, which are then transferred to each VMM3a ASIC via an add-on flex PCB (see Sec. \ref{sec:card_flex}). It receives two input signals via micro-Coaxial connectors and outputs identical copies to three Hybrids. Controlled-impedance traces matched to 50 $\Omega$ were used throughout to preserve signal integrity. The choice of a separate add-on module enabled testing of various buffer designs without redesigning the entire PCB.

On top of the PCB, six 140-pin card connectors are placed, to which the adapter cards of the Hybrids are connected (see Sec. \ref{sec:card_flex}). The channels are routed such that for each Hybrid, 112 of the 128 available channels are dedicated to signals from the TPC pad plane, with 16 spare channels. For the latter, eight (four per VMM3a ASIC) are used for synchronization signals.

The pad-plane adapter assembly was performed manually. Since no Ball Grid Array (BGA)-like connectors of this specification are commercially available, THT connectors were adapted for use as surface-mount components. For the soldering process, the components on the bottom side were first assembled, followed by those on the top side. Soldering the 2-mm-pitch connectors required a dedicated assembly procedure. A bismuth–tin solder paste with a low melting point of 138$^\circ$C was applied, while the edge pins were fixed in place using lead-free tin solder of higher melting point to prevent the connectors from shifting during reflow. Reflow soldering was performed on a heat plate with a 10 kg weight applied to the PCB to keep the connectors straight and compensate for fabrication tolerances by slightly deforming them under heat.

For mounting the connectors on the top side, the standard Surface-Mount Device (SMD) reflow technique was applied. However, during this step, heat was transferred from the hot plate to the already-mounted 2-mm-pitch connectors. To prevent misalignment under thermal stress, again, a 10 kg weight was placed on top of the card connectors to maintain contact and compensate for any PCB deformation. After soldering, the fully assembled PCB was cooled gradually over 1.5 hours to minimize mechanical stress and reduce the risk of connector delamination.

\subsubsection{Modification of Hybrids}

\begin{figure}[ht]
\includegraphics[width=\textwidth]{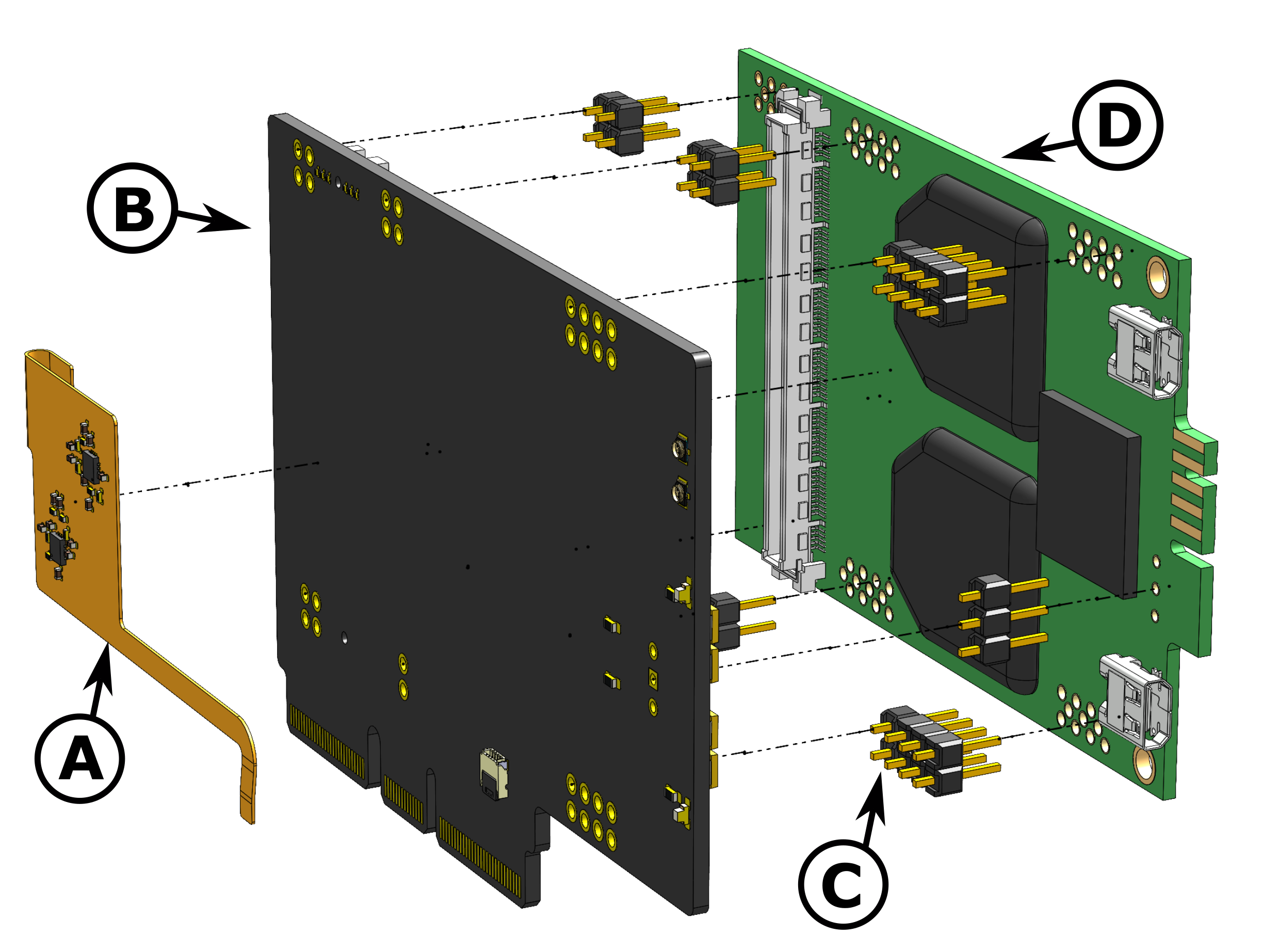}
\centering
\caption{Exploded view of the modified VMM3a Hybrid assembly: (A) flexible PCB add-on; (B) card adapter PCB; (C) 2.54 mm pitch pin headers; and (D) modified VMM3a Hybrid (power connector and heatsinks removed).}  
\label{figure: Hybrid modifications}
\end{figure}

The standard VMM Hybrid board exceeded the spatial constraints of the HYDRA TPC setup, requiring physical modifications to its baseline configuration. In particular, to fit all boards on top of the TPC pad plane, the thickness of the Hybrid ($\sim$15 mm) had to be reduced. For this purpose, several components were removed from the Hybrids: the front and back heat sinks and their connectors, as well as the power connector. These were replaced by 2.54 mm pin headers, as shown in Fig. \ref{figure: Hybrid modifications}. Following these modifications, an alternative cooling solution for the FEE is needed. For measurements presented in this paper, fans were used, while a dedicated cooling system for the operation inside the GLAD magnet is being developed.

The Hybrids are usually mounted horizontally, with the 140-pin Hirose FX10A connector mated directly to its counterpart on the detector. In the HYDRA setup, however, the modules are installed vertically. The connection to the detector is provided via a dedicated card adapter, which establishes electrical interfaces for the signal channels, the power supply, and the synchronization signals via a flex add-on. These modifications successfully reduced the board's thickness by nearly 50\%, bringing it down to 8.5 mm.

\subsubsection{Card adapter and Flex add-on}\label{sec:card_flex}
\label{Card adapter}
\begin{figure}[ht]
\includegraphics[width=\textwidth]{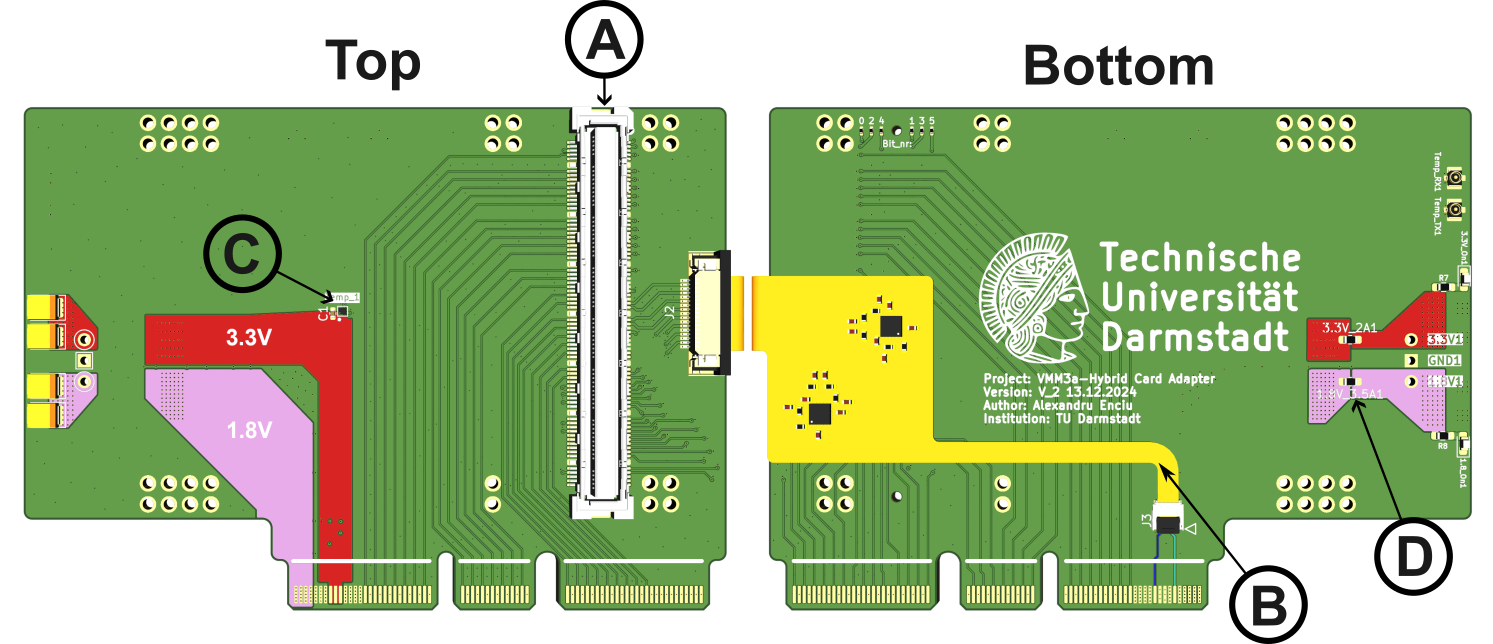}
\centering
\caption{Card adapter and flex add-on PCBs: (A) Hirose FX10-140S connector; (B) Flex add-on PCB; (C) 1-wire temperature sensor TMP104; (D) Over-current fuses of 2 and 3.5 A for the 3.3 V and 1.8 V power rails.}
\label{figure: Card Adapter}
\end{figure}

A card adapter PCB (see Fig.~\ref{figure: Card Adapter}) is used to accommodate the modified VMM3a Hybrid and to provide electrical connections between the Hybrid and the TPC. This board is an 8-layer FR4 PCB with blind and buried vias, with the smallest trace width of 0.1 mm. To reduce crosstalk among the high-density routed traces, a via-stitched coplanar ground plane is placed between them. At the VMM3a ASIC position, a temperature sensor controlled digitally via a 1-Wire protocol is used to independently monitor the temperature. The two clock signals used for synchronization are delivered to the VMM3a ASICs via an add-on flexible 2-layer PCB. This flex circuit actively splits and buffers the two incoming signals using two clock buffer integrated circuits, yielding eight separate signals. These are then AC-coupled to the ASIC's input channels via 500 fF capacitors.

\subsection{Event selection and synchronization}\label{sec:sync}

Synchronization of the HYDRA TPC readout with the R$^3$B DAQ system is a fundamental requirement. 
 The R$^3$B DAQ includes various so-called DAQ nodes, corresponding to the different detection systems of the R$^3$B setup. Each node emits data with timestamps that enable correlation of events across separate DAQ groups. At GSI-FAIR, timestamps originate from a master timekeeper and are distributed via the White Rabbit (WR) protocol to all timestamp receivers. It is an Ethernet-based network for clock and time distribution for a large number of nodes with nanosecond accuracy \cite{Beck2012}. Consequently, each DAQ group must contain such a receiver. However, the VMM3a/SRS readout lacks native support for a timestamp receiver, and therefore direct integration into the R$^3$B timing scheme is not possible.

\begin{figure}[ht]
\includegraphics[width=0.9\textwidth]{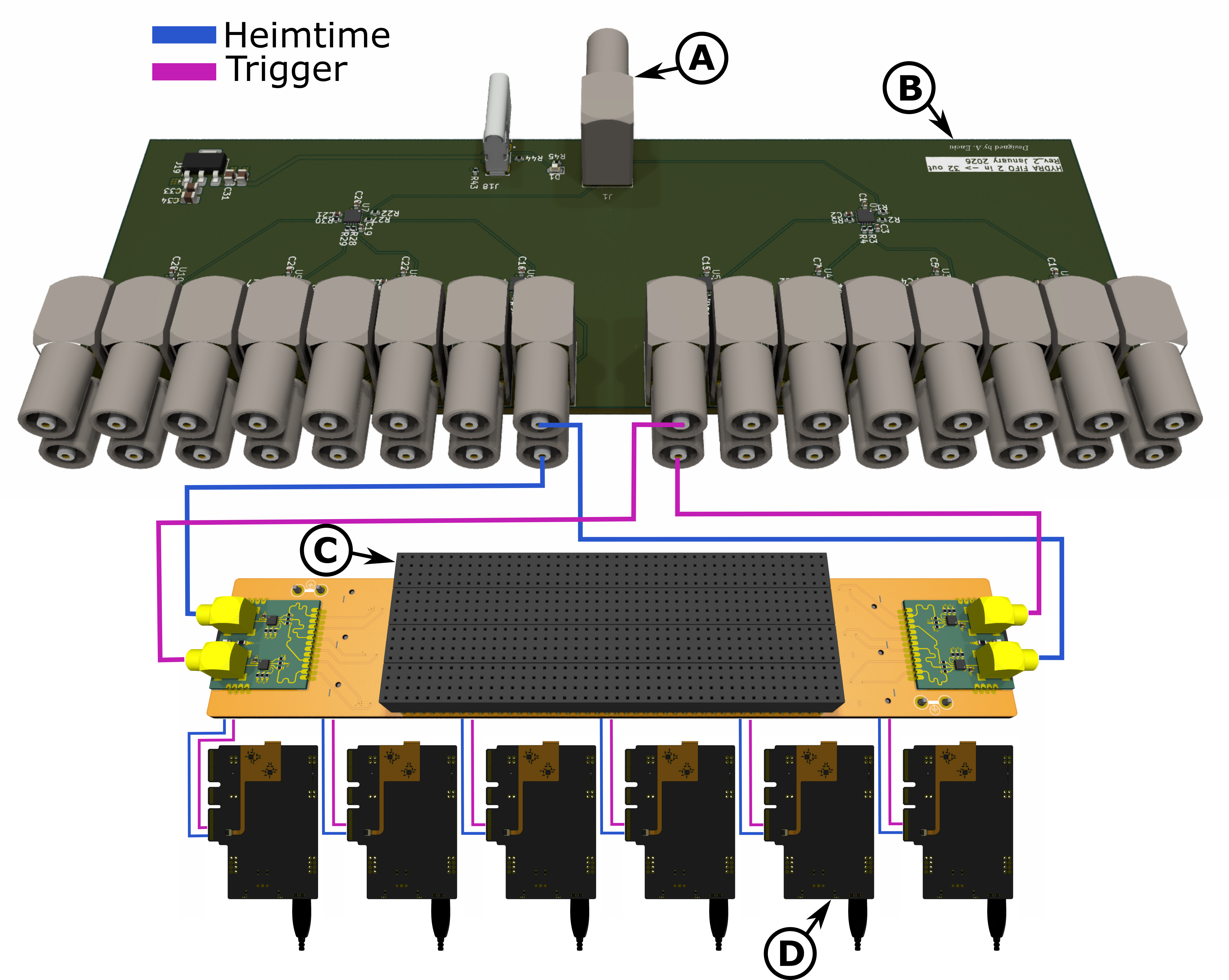}
\centering
\caption{Distribution of the synchronization signals into the HYDRA-TPC readout. The Heimtime and trigger signals are sent from the VULOM to a coaxial connector (A) of the distribution board (B) with 16 signal pairs, which connect to the 8 pad-plane adapters (C). In the adapter, they are further split and routed using add-on clock buffers and flex PCBs into each VMM3a card (D).}
\label{figure: Clock distribution logic}
\end{figure}

For such cases where the DAQ system cannot use the serial time protocol, the heimtime protocol was implemented \cite{TRLO}. Given that the system can locally timestamp an input logical signal, it can receive the periodic signals of the heimtime protocol. Then, the global timescale can be decoded from the data during online processing. The heimtime signal is generated by the VULOM (VME Universal Logic Module) \cite{VULOM} and distributed to dedicated channels in each VMM3a ASIC along with a trigger signal. The VULOM module also produces the trigger logic and sends out a trigger signal when a trigger generated by $e.g.$ the scintillator wall is accepted. By verifying the trigger signal's length and recording its leading and trailing edges, time synchronization across different detection systems can be validated.

The implementation of the heimtime and trigger signal into the HYDRA readout is illustrated in Fig. \ref{figure: Clock distribution logic}. First, they are delivered as 16 signal pairs to both sides of the eight pad-plane adapters by a time-trigger distribution board. On each side, they are fanned out by the add-on clock buffers with an I/O ratio of 1:3 to the card adapters via the add-on flex PCB. On each card adapter, they are further split by dedicated buffers with an I/O ratio of 1:4, and two outputs per buffer are used to feed the two VMM3a ASICs on the corresponding Hybrid. Although one output per signal is sufficient, one additional one is used as backup. The synchronization signals are AC-coupled into the spare channels of the VMM3a ASICs via 500 fF capacitors.

\section{Performance evaluation}

\subsection{Crosstalk evaluation}

The card adapter routes both detector signals and the external heimtime and trigger signals. Eight spare channels on each VMM Hybrid are assigned to these 3.3~V timing signals. Their logic transitions are the largest voltage excursions routed on the adapter and were therefore selected as the worst-case aggressor for the crosstalk estimate. The simulation evaluates deterministic coupling from a timing-signal trace into an adjacent detector-channel trace. It is not used to determine the electronic noise, which is measured separately using the S-curve method.

A representative pair of adjacent traces was modeled in LTspice \cite{ltspice}. The traces have a width of $0.1~\mathrm{mm}$ and a modeled length of $l=60~\mathrm{mm}$. Their center-to-center spacing is $350~\upmu\mathrm{m}$, corresponding to 3.5 times the trace width, and via-stitched ground copper is placed between them \cite{Welch2018}. Together with the reference plane, this arrangement forms a grounded coplanar routing geometry \cite{wadell1991transmission}.

The self-capacitance and self-inductance of each trace, as well as the mutual capacitance and inductance between the two traces, were calculated from the geometry and dielectric properties of the eight-layer FR4 stackup \cite{bogatin2010signal}. The resulting per-unit-length parameters included the mutual capacitance $C'_m$ and mutual inductance $L'_m$. These values were multiplied by the modeled trace length to obtain the lumped parameters $C_m=C'_m l$ and $L_m=L'_m l$ used in the circuit simulation \cite{Cheng2021}.

\begin{figure}[ht]
    \centering
    \includegraphics[width=\textwidth]{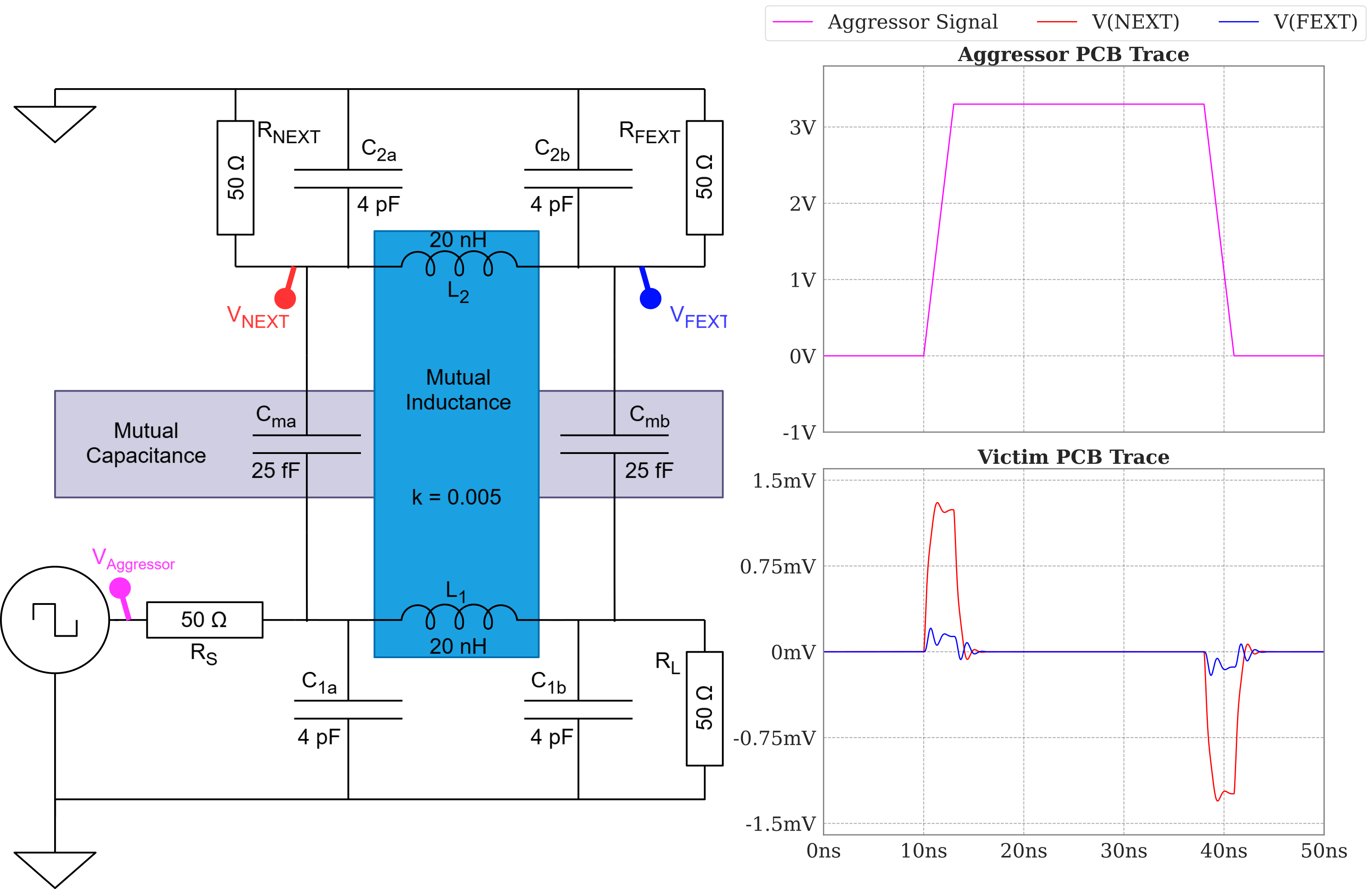}
    \caption{LTspice model and transient response used to estimate coupling from a 3.3~V timing-signal trace into an adjacent detector-channel trace. Left: lumped-element model and probe locations for the aggressor voltage ($V_{\mathrm{Aggressor}}$), near-end crosstalk ($V_{\mathrm{NEXT}}$), and far-end crosstalk ($V_{\mathrm{FEXT}}$). Right: simulated aggressor, NEXT, and FEXT waveforms for a 3~ns input edge.}
    \label{figure: cross_talk_simulation}
\end{figure}

Each trace was represented by a lumped-element $\pi$-network \cite{johnson1993high}, as shown in Fig.~\ref{figure: cross_talk_simulation}. The total self-inductance of the aggressor trace was represented by the series element $L_1$, while its self-capacitance to ground was divided equally between $C_{1a}$ and $C_{1b}$. The victim trace was represented in the same way by $L_2$, $C_{2a}$, and $C_{2b}$.

Capacitive coupling was included by dividing the total mutual capacitance between the near and far ends of the two $\pi$-networks \cite{Zhang2012}:

\begin{equation}
    \label{equation: Capacitive_coupling}
    C_m=C_{ma}+C_{mb}.
\end{equation}

Inductive coupling was introduced through the mutual inductance $L_m$ and the coupling coefficient $k$:

\begin{equation}
    \label{equation: Inductive_coupling}
    L_m=k\sqrt{L_1L_2}.
\end{equation}

The coupling coefficient entered in the LTspice model was therefore

\begin{equation}
    \label{equation: Coupling_factor_k}
    k=\frac{L_m}{\sqrt{L_1L_2}}.
\end{equation}

Because the two traces have the same geometry, $L_1=L_2$, and Eq.~\ref{equation: Coupling_factor_k} reduces to $k=L_m/L_1$.

The aggressor trace was driven by a 3.3~V square pulse with rise and fall times of 3~ns, representing the fastest expected transition of the timing-distribution signals. The 3.3~V amplitude therefore represents a timing signal rather than a detector pulse. The aggressor was terminated by a $50~\Omega$ source resistance $R_S$ and a $50~\Omega$ load resistance $R_L$. The near and far ends of the victim trace were terminated by $R_{\mathrm{NEXT}}=50~\Omega$ and $R_{\mathrm{FEXT}}=50~\Omega$, respectively. These values approximately match the calculated characteristic impedance,

\begin{equation}
    Z_0 \approx
    \sqrt{\frac{L_1}{C_{1a}+C_{1b}}},
\end{equation}

and provide defined boundary conditions while limiting reflections in the simulated trace pair.

The transient response is shown in Fig.~\ref{figure: cross_talk_simulation}. The peak near-end crosstalk, measured across $R_{\mathrm{NEXT}}$, was approximately 1.3~mV, corresponding to 0.039\% of the 3.3~V aggressor transition. The far-end transient, measured across $R_{\mathrm{FEXT}}$, remained below 0.2~mV, corresponding to less than 0.006\%. For the modeled geometry and termination conditions, coupling from the timing-signal trace into the adjacent detector-channel trace therefore remained below 0.1\%.

The calculation supports the selected trace spacing, which satisfies the 3W design rule \cite{Mudavath2019}, together with the use of via-stitched grounded copper between adjacent traces. The simulated transients represent deterministic trace-to-trace coupling and should not be interpreted as electronic noise or equivalent noise charge. The noise performance of the assembled readout is evaluated independently from the S-curve and ENC measurements presented in the following subsection.

\subsection{Electronics noise and S-curve}

An important aspect in the characterization of the system is the estimation of the electronic noise affecting the detector signals. The Equivalent Noise Charge (ENC) is typically used to quantify the noise level. The electronic noise of the VMM has been investigated previously $e.g.$ in~\cite{Geronimo2022,terzimpasoglou2021}, as a function of the electronics gain and peaking time. In these studies, the noise was measured using an analogue monitoring output of the VMM and sampling the waveforms with an oscilloscope. As these studies rely on the analogue monitoring output and external devices, here we follow the approach proposed in~\cite{scharenberg2023}, which uses the readout system solely. 

\begin{figure}[t]
\includegraphics[width=\textwidth]{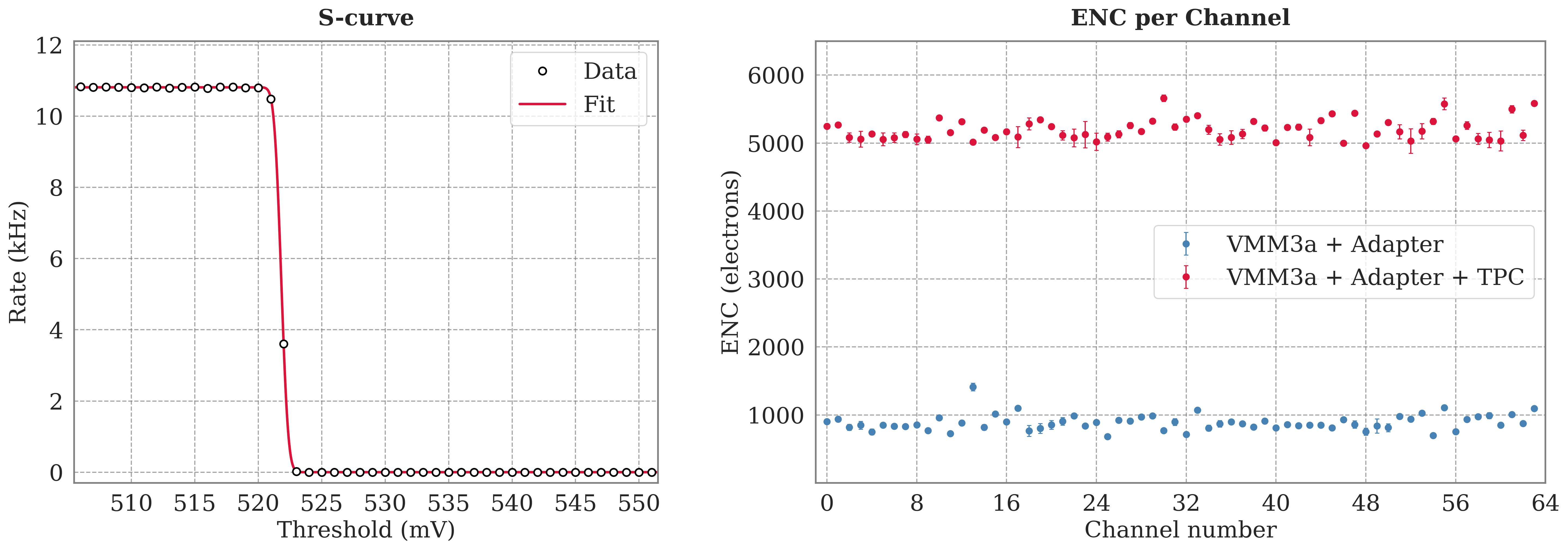}
\centering
\caption{Left: Example of S-curve measurement, number of counts per threshold level for an internal pulse signal in one VMM3a channel together with a fit following Eq.~\ref{eq:noise}. Right: Electronic noise for all channels in one VMM3a chip w/o (blue) and w/ (red) the TPC. The measurement was performed with a peaking time of 200 ns and electronics gain of 3 mV/fC.}
\label{figure: S_curve}
\end{figure}

It is based on the S-curve method~\cite{kolanoski2020}, where an internal pulser generates signals with a fixed amplitude that are injected into the VMM readout channels while increasing the threshold level. In the absence of noise, when the threshold is below the pulse amplitude, all signals would be recorded, with no signals when it exceeds the pulse height. In practice, due to the electronic noise, the output height of the preamplifier is broadened, leading to a so called S-curve. An example for an S-curve measurement of one VMM3a channel is shown in the left panel of Fig.~\ref{figure: S_curve}. The noise ($\sigma_{\rm noise}$) is extracted by fitting the data an error function
\begin{equation}
    N(x) = \frac{N_{\rm max}}{2} {\rm erfc} \left ( \frac{x-x_{50}}{\sqrt{2}\sigma_{\rm noise}} \right ),
    \label{eq:noise}
\end{equation}
where $N_{\rm max}$ is the maximum number of counts, $x$ is the threshold level, and $x_{50}$ the threshold level where 50\% of the maximum is reached. This way, the width $\sigma_{\rm noise}$ is extracted for each VMM3a channel, and can be linked directly to the ENC via
\begin{equation}
    {\rm ENC} = 6242 \frac{\sigma_{\rm noise}}{G_{\rm elec}},
\end{equation}
where $G_{\rm elec}$ is the electronics gain, and 6242 the number of electrons per 1~fC. The right panel of Fig.~\ref{figure: S_curve} shows the ENC for all channels in one VMM3a chip w/o (blue) and w/ (red) the TPC. Without the detector being on, the electronics noise is around 1000 electrons, which is consistent with the results obtained in Ref. \cite{scharenberg2023} where no adapter boards were used or modifications to the Hybrids were done. In contrast, ramping up the high voltage applied to the TPC increases the measured ENC by a factor of 5 (to 5000 electrons). This demonstrates that the detector itself is the dominant source of noise, whereas the modified readout architecture successfully maintains the baseline noise level.

\subsection{Cosmic-rays - time correlation with scintillator wall}

For the cosmic ray measurements, the TPC was equipped 50\% the readout chain described in Sec. \ref{sec:vmm}, $i.e.$, 4 complete pad-plane adapters with 24 Hybrids. An electronics gain of 3~mV/fC and a peaking time of 200~ns were used for the VMM3 settings. To validate the performance of the system and the time synchronization, measurements were performed using cosmic muons, where the scintillator wall was attached to the back side of the TPC. The TPC was operated with an Ar-CO$_2$ (90-10) gas mixture and a gain of $\mathcal{O}(10^4)$. The wall consists of 16 plastic scintillator bars each with a length of 250 mm and width of 23~mm~\cite{Ji2026}. It provides the trigger to the DAQ, set when at least one bar has a signal above the threshold. 

\begin{figure}[t]
\includegraphics[width=\textwidth]{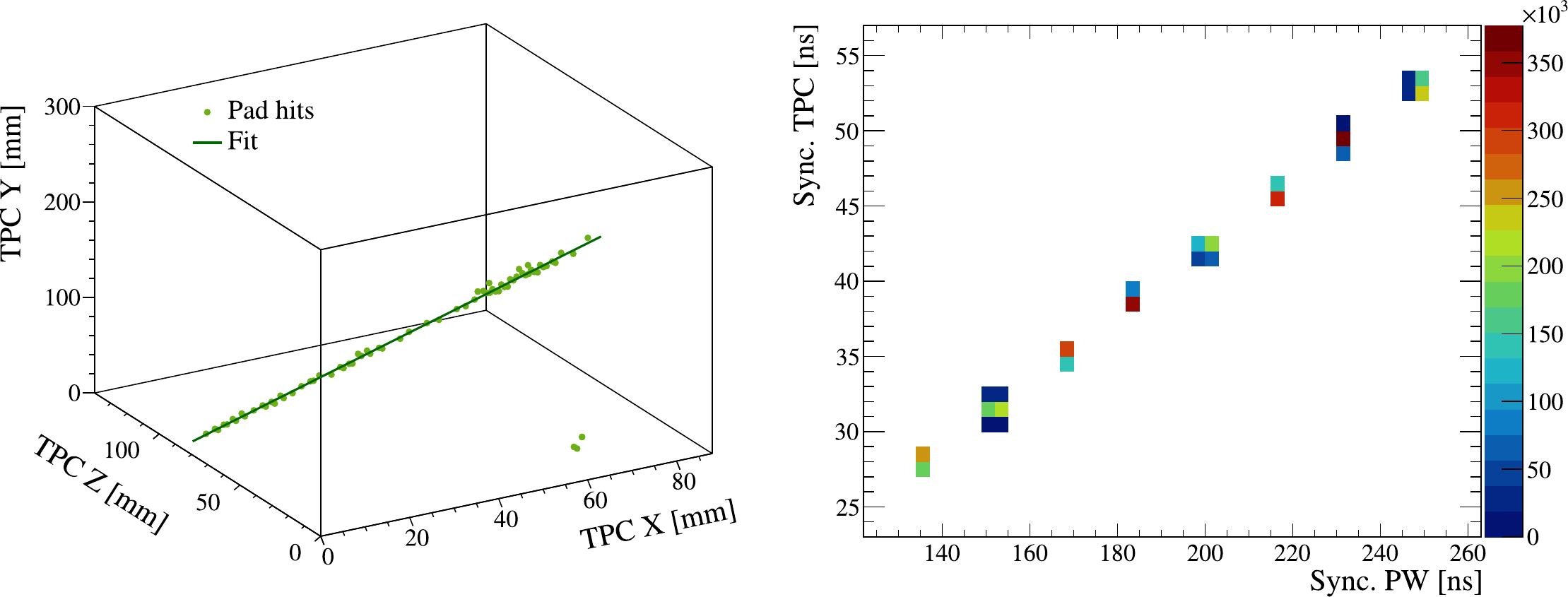}
\centering
\caption{Left: Example of 3D track reconstruction for a cosmic muon event in the TPC with a trigger from the scintillator wall. The solid line represents a linear fit to the hit points. Right: Synchronization check between the TPC and plastic scintillator wall (PW) obtained from the trigger signal recorded in both systems. } 
\label{figure: TPC_PW}
\end{figure}

With this trigger condition, muon tracks were reconstructed in the TPC. The left panel of Fig.~\ref{figure: TPC_PW} shows an example 3D reconstruction of such an event together with a linear fit. The transversal coordinates (X,Z) for each point are determined from the hit position on the pad plane. The longitudinal one is determined from the measured drift time and the known drift velocity in the gas. The drift time is extracted relative to the start time provided by the scintillator wall. As the data from both systems are merged, it is important to ensure that the two are synchronized. This can be achieved by comparing the trigger signal discussed in Sec.~\ref{sec:sync} recorded by the TPC and scintillator wall, presented in the right panel of Fig.~\ref{figure: TPC_PW}. Shown are events with a valid trigger from the scintillator wall that have non-zero signals in the TPC. A clear correlation can be seen, with no off-diagonal events, verifying a reliable synchronization.

\section{Conclusion}

In this work, we have presented a compact implementation of the high-rate VMM3a-based readout system for the HYDRA time projection chamber. The integration of the VMM3a readout posed significant challenges due to the requirement for a compact design and the high channel density of the TPC pad plane (25 channels/cm$^2$). These challenges were successfully addressed through dedicated modifications of the front-end electronics and the development of custom adapter boards, allowing the installation of 48 VMM3a Hybrids in a vertical configuration directly on top of the detector. The resulting system achieves the required channel density while maintaining excellent signal integrity. Measurements demonstrate that crosstalk and equivalent noise charge remain at low levels, with the modified Hybrids exhibiting performance comparable to that of unmodified modules.
Another challenge was synchronizing the VMM3a readout with other detector systems, as the VMM3a architecture does not provide native support for a timestamp receiver. This limitation was overcome by distributing timing and trigger signals to dedicated ASIC channels, enabling timestamp reconstruction from the recorded data and reliable verification of synchronization. The implemented timestamp-injection scheme is fully scalable and has been integrated into both the adapter card and pad-plane adapter architecture through dedicated add-ons.

The complete readout chain was validated in operation with the HYDRA-TPC using cosmic-ray measurements triggered by a plastic scintillator wall. The successful reconstruction of particle tracks and the observed time correlation between the TPC and scintillator signals demonstrate the correct functionality of the detector-readout chain. These results establish the developed VMM3a-based readout as a robust and scalable solution for high-density, high-rate TPC applications, such as the HYDRA TPC at R$^3$B, GSI/FAIR at the origin of this work.

\section*{Acknowledgements}
    \noindent This work was supported by the Helmholtz Forschungsakademie Hessen f{\"u}r FAIR, Germany, the German Federal Ministry of Education and Research - BMBF project numbers 05P21RDFNB and 05P24RD1, the State of Hesse within the Research Cluster ELEMENTS (Project ID 500/10.006), as well as the Alexander von Humboldt foundation.

\bibliographystyle{unsrt}
\bibliography{bib}

\end{document}